\documentclass[aps,preprint,onecolumn]{revtex4}
\usepackage{graphicx}

\begin{document}
\bibliographystyle{prsty}
\title{Diffraction by a small aperture in conical geometry: Application to  metal coated tips used in near-field scanning optical microscopy}
\author{A.~Drezet, J.~C.~Woehl, and S.~Huant}
\address{Laboratoire
de Spectrométrie Physique, CNRS UMR5588, Universit\'e Joseph Fourier Grenoble,
BP 87, 38402 Saint Martin d'Hères cedex, France }

\begin{abstract}
Light diffraction through a subwavelength aperture located at the apex of a metallic screen with conical geometry is investigated theoretically. A method  based on a multipole field expansion is developed to solve Maxwell's equations analytically using boundary conditions adapted both for the conical geometry and for the finite conductivity of a real metal. The topological properties of the diffracted field are discussed in detail and compared to those of the field diffracted through a small aperture in a flat screen, i.~e. the Bethe problem. The model is applied to coated, conically tapered optical fiber tips that are used in Near-Field Scanning Optical Microscopy. It is demonstrated that such tips behave over a large portion of space like a simple combination of two effective  dipoles located in the apex plane (an electric dipole and a magnetic dipole parallel to the incident fields at the apex) whose exact expressions are determined. However, the large ``backward'' emission in the P plane - a salient experimental fact that remained unexplained so far - is recovered in our analysis which goes beyond the two-dipole approximation.
\end{abstract}
\newpage
%
\maketitle
\section{Diffraction and near-field optical microscopy  }
Diffraction and scattering are among the most difficult phenomena encountered in optics. Eigenvalue problems with complex boundary conditions have to be solved in order to account for the finite wavelength of light in the interactions between optical waves and matter. Exact solutions are extremely rare and limited to simple, idealized diffracting objects.\\
In general, approximation methods are used for diffraction problems involving apertures or obstacles of arbitrary geometry.
For example, the interaction of light with a screen in the limit where the wavelength is short compared to the dimensions of the obstacles was already treated a long time ago by Huygens, Fresnel, and Kirchhoff (among others). This case corresponds to small deviations from geometrical optics. However, it is totally irrelevant for phenomena implying strong near-field patterns which occur for wavelengths comparable to the dimensions of the diffracting object. In these cases, the interaction of light with matter cannot be considered as a small perturbation, and the related boundary problem must be solved exactly.\\

This is the situation encountered in Near-Field Scanning Optical Microscopy (NSOM) in which strong interactions exist between a nanosource and nearby nano-objects\cite{VanLabeke,Girard,Greffet,Novotny2}.
In NSOM, the specimen is usually illuminated through a subwavelength aperture located at the apex of a coated, tapered fiber tip. Such an optical tip can be mimicked by a truncated metallic cone, i.~e. a very complex geometry. The aim of the present paper is to study this diffraction geometry in order to elucidate the properties of actual tapered optical tips. Although a rigorous solution of the problem will not be given, we will make use of an approximation method based on a field expansion in "quasi" spherical multipoles which is valid in the far-field. Maxwell's equations are solved analytically within this scheme, which allows us to describe in detail the topological properties of the diffracted field. In particular, we will show that the dipole behavior anticipated earlier\cite{Karraia,Karraib} can be formally justified and that it remains valid over a large portion of space. The predictions of our analysis are found  to be in a fair agreement with available far-field emission patterns of actual conical optical tips.\\

The paper is organized as follows. In section II, the problem of diffraction through an aperture in a metallic screen is introduced in general terms. In section III, a method to solve electromagnetic problems with conical geometry is developed. The radiation pattern of a conical optical probe is described in section IV for the case of a tip coated with a perfect metal, which is extended to real metals in section V. Section VI compares the theoretical results with experimental emission patters of metal coated, optical fiber tips. A summary is given in section VII.
%
\section{Introduction to the problem of diffraction by small apertures : The Bethe-Bouwkamp solution }
\begin{figure}[h]
\begin{center}
\includegraphics[width=10cm]{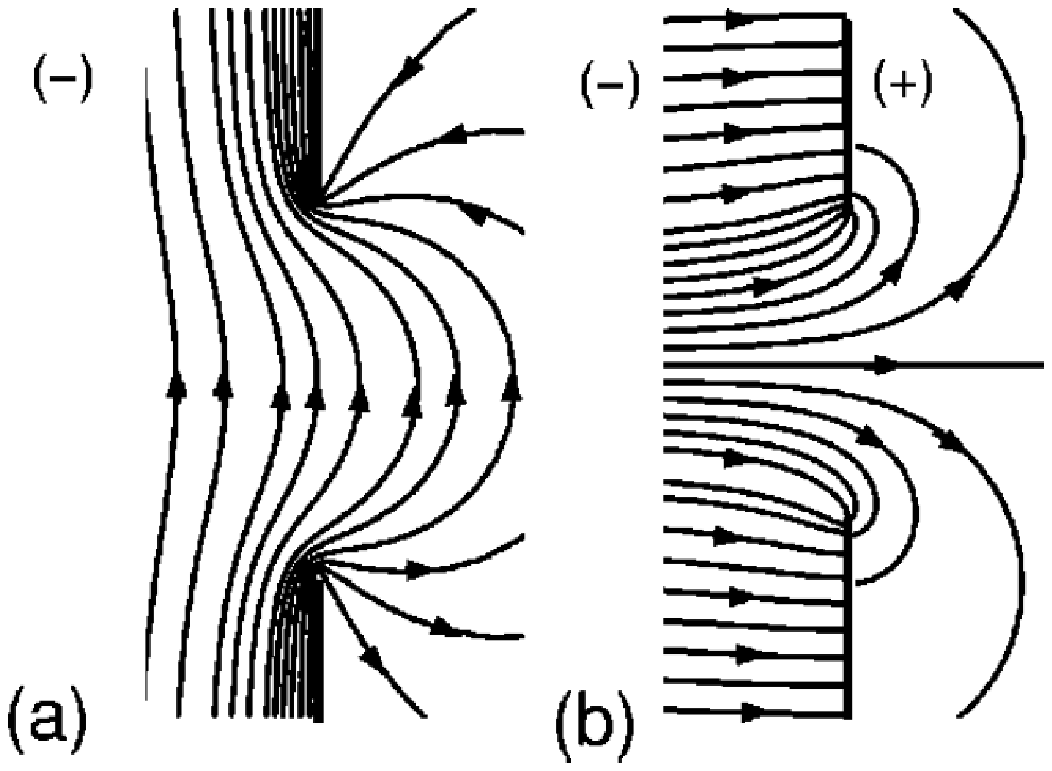}
\caption{A rigourous representation of the magnetic (a) and electric (b) fields in the vicinity of a circular aperture in a metallic screen. The aperture has subwavelength size and the screen has an infinite conductivity. The fields reduce to uniform far-fields ${\mathbf E}_{0}$, ${\mathbf B}_{0}$ incident part of space (-), and to a dipolar field in the other part (+).  }
\label{bethe}
\end{center}
\end{figure}

 In classical optics, the diffraction of light by an aperture in a screen is studied within the Kirchhoff approximation \cite{Kirchhoff,Poincare}. The use of this scalar method is justified in numerous cases where the polarization of light can be ignored and where the aperture size is large compared to the wavelength of the incident radiation. In order to apply the Huygens-Fresnel principle to light diffraction by small apertures, Smythe\cite{Smythe,Huang} has developed a vectorial Kirchhoff integral formula based on the Green function method.
The formalism is very general but - like Kirchhoff's scalar theory - its practical application is restricted to long wavelengths compared to the geometrical parameters of the aperture. The first-order solution consists in neglecting the influence of boundary effects on the light in the aperture zone. This method, which may be viewed as the electromagnetic counterpart of the Born\cite{Hecht,Born} approximation in quantum mechanics, is adapted to the quasi geometrical regime of Maxwell's equations. However, it is not adapted to long wavelength radiation.\\ Bethe \cite{BetheHans} (and after him Bouwkamp \cite{Bouwkampa}) have developed a rigorous solution for the case of a small circular aperture in a perfect metallic plane by solving an electrostatic and a magnetostatic eigenvalue problem for the static near-field existing in the vicinity of the aperture. This solution, which generalizes previous work by Rayleigh \cite{Rayleigh}, can be easily recovered by using oblate spheroidal coordinates and harmonical wavefunctions. The Helmoltz equation
 $[{\mathbf \nabla}^{2}+k^{2}]\psi=0$, which reduces to the Laplace equation for small hole sizes  ${\mathbf \nabla}^{2}\psi=0$, has the solution
$\psi\left(\xi,\eta,\phi \right)=P^{m}_{l}\left(\xi\right)\cdot P^{m}_{l}\left(i\eta\right).e^{\pm im\phi}$, where
$P^{m}_{l}$ denote Legendre wavefunctions indexed by two integers $l,m$. $\xi,\eta,\phi$ are the oblate spheroidal coordinates. The Bethe solution exhibits a	complicated behavior in the vicinity of the aperture. Nevertheless, the Bethe solution reduces to a multipole expansion in the near-field domain limited by $r/\lambda\sim 1$ (which corresponds to $r\rightarrow +\infty$ for the static Laplace equation where $\lambda\rightarrow +\infty$). In the particular case of an incident plane wave, it reduces to a electromagnetic dipole field (see Fig.~\ref{bethe}). The corresponding effective dipoles ${\mathbf P}_{\textrm{eff}}$ and
 ${\mathbf M}_{\textrm{eff}}$ depend on the aperture radius and on the incident electromagnetic field ${\mathbf E}_{0}$, ${\mathbf B}_{0}$
as given by the formula ${\mathbf P}_{\textrm{eff}}=\mp\frac{1}{3\pi}a^{3}{\mathbf E}_{0,\perp}$,
${\mathbf M}_{\textrm{eff}}=\pm\frac{2}{3\pi}a^{3}{\mathbf B}_{0,\|}$. The signs in front of the expressions refer to the ``incident + reflected'' domain and to the ``transmitted'' domain, respectively. It is interesting to note that the factor of 2 between  ${\mathbf M}_{\textrm{eff}}$ and ${\mathbf P}_{\textrm{eff}}$ appears in all similar situations as we will see throughout this paper.
 The physical signification of the Bethe dipoles can be understood using the Clausius Mossoti\cite{Jackson} formula. It describes the polarization produced by a locally constant electromagnetic field ${\mathbf E}_{0}$, ${\mathbf B}_{0}$ in a small dielectric sphere of constant permittivities $\epsilon,\mu$ which is immersed in a homogenous medium with permittivities $\epsilon_{0},\mu_{0}$. According to this formula, we have ${\mathbf P}=\epsilon_{0}\frac{\epsilon/\epsilon_{0}-1}{\epsilon/\epsilon_{0}+2}a^{3}{\mathbf E}_{0}$, and ${\mathbf M}=\frac{1}{\mu_{0}}\frac{\mu/\mu_{0}-1}{\mu/\mu_{0}+2}a^{3}{\mathbf B}_{0}$. With the condition  $\epsilon/\epsilon_{0}, \mu_{0}/\mu\ll 1$ for a hole in a perfect metal, we obtain the following two relations: ${\mathbf P}=-\frac{\epsilon_{0}}{2}a^{3}{\mathbf E}_{0}$, ${\mathbf M}=\frac{1}{\mu_{0}}a^{3}{\mathbf B}_{0}$ which are related by the same factor of 2 as for the Bethe case. \\
The diffraction of light by a small aperture in a perfectly conducting plane can be regarded as the complementary case to the scattering of a wave by a small conducting particle.
This point of view is in agreement with the electromagnetic Babinet\cite{Babinet} theorem which yields this result  directly ( see Smythe \cite{Smythe}, Landau \cite{Landau}).  Rayleigh has developed the theory of diffraction by small particles and has found that the shape of the particle is not a fundamental parameter if one is interested in the far-field pattern of the scattering wave only. This is due to the fact that the light cannot distinguish the shape of the particle if the wavelength is very large compared to $a$. Therefore, we anticipate that the existence of two effective dipoles and their far-field behavior is very general and does not depend on the shape of the small hole or the screen geometries. The exact expressions of these dipoles may depend on the hole shape and their directions may differ from the Bethe case, but the factor of two between $M$ and $P$ as well as the presence of the fields and their dependance on the third power of the radius constitute probably a general result. A validity criterion for this statement can be found in the curvature of the surface at the hole location. If the curvature is too strong (i.~e. if the curvature radius is too small), the hypothesis cannot be considered as reasonable due to the presence of strong boundary effects at the surface.  Therefore,  we expect the validity of the dipole model to diminish with the curvature radius. More precisely, the spatial domain in which the dipole model is applicable reduces progressively to a small solid angle around the optical axis when the curvature radius decreases, i.~e. when boundary effects invade space. We will see below that these general trends are confirmed by theory.

%
%
\section{The quasi-multipole method }
The method of quasi-multipoles developed here is the result of a convergence of three domains in electromagnetism. It lies at the intersection of i) the metallic wave guide theory for guides with variable cross sections \cite{Marcuvitz,Collin,Butler}, ii) the spherical multipole expansion method developed by Bouwkamp\cite{Jackson,Bouwkampb}  for electromagnetic fields applicable to localized source distributions, and iii) the Hall\cite{Hall} equilibrium solutions for conductors with conical geometry.
For instance, the electromagnetic field radiated by a conical antenna or a conical hole can be described using this formalism.
In order to simplify the discussion, we will use the same notations for spherical multipoles as those presented by  Jackson\cite{Jackson}. Let us consider a source-free region of space with permittivities $\epsilon $ and $\mu$. A conical surface with half-angle $\beta$ separates this space into an ``oustide'' region and an ``inside'' region. If $\beta$ exceeds $\pi/2$, the external and the internal regions are mutually exchanged. For simplicity, these zones are considered as decoupled and totally independent. This is possible in the case of a perfectly conducting metallic surface in which tunneling of the field through the surface is forbidden. From now on, we will only consider the external problem. Maxwell's equations can be written (assuming an $e^{-i\omega t}$ time dependence):
\begin{eqnarray}
{\mathbf E}=\frac{i}{k\epsilon}{\mathbf \nabla \times H} & \quad;\quad &{\mathbf H}=-\frac{i}{k\mu}{\mathbf \nabla \times E} \nonumber \\{\mathbf \nabla}.{\mathbf E}=0 &\quad;\quad & {\mathbf \nabla}.{\mathbf H}=0.
\end{eqnarray}
Making use of the identity $\nabla^{2}\left({\mathbf r.A}\right)={\mathbf r}\cdot\left( \nabla^{2} {\mathbf A}\right)+2{\mathbf \nabla\cdot A}$, these equations reduce to
\begin{eqnarray}
\left( \nabla^{2} + k^{2}\epsilon \mu \right){\mathbf r.E}=0 &\quad;\quad& \left( \nabla^{2} + k^{2}\epsilon \mu \right){\mathbf r.H}=0.
\label{nabla}
\end{eqnarray}
 By writing the Laplace operator in spherical coordinates, and separating the angular and radial variables, a typical solution of Eq.~\ref{nabla} can be expressed as follows:
\begin{equation}
\psi\left({\mathbf r}\right)=\sum_{\nu ,m,\pm} \alpha_{\nu,m}^{\pm} h_{\nu}^{\pm}\left(k\sqrt{\epsilon\mu}r\right)Y_{\nu,m}\left(\theta,\phi\right).\label{somme}
\end{equation}
Both Hankel functions $h_{\nu}^{\pm}$ appear in this expansion ( the $\pm$ signs refer to outgoing and incoming waves, respectively). They satisfy the spherical Bessel equations for the radial variable r and the quasi-harmonic functions $Y_{\nu, m}\propto P^{m}_{\nu}\left(\cos{\theta}\right)\cdot e^{im\phi}$ defined in the literature\cite{Hall,Bateman,Abramowitz}. The latter represent the generalization to a conical geometry\cite{Hall} of the well-known spherical harmonics $Y_{l,m}$ used in spherical eigenvalue problems.
However, an important modification with respect to spherical harmonics is the presence of a non-integer parameter $\nu$ which replaces the integer $l$ and which depends on the boundary conditions on the cone. The possible values of $\nu$ in Eq.~\ref{somme} follow an infinite sequence which is a function of $\beta$: $\nu_{p}=\nu_{p}\left(\beta\right)$ ($p$ integer). The exact relation depends on the boundary conditions on the cone.\\
In analogy with the spherical case and with wave guide theory, we now decompose the electromagnetic field obeying a specific boundary condition on the cone into two parts: a magnetic quasi-multipole part (M) of the transverse electric field (TE), and an electric quasi-multipole part (E) of the transverse magnetic field (TM). These two parts are characterized by components of order ($\nu,m$) satisfying
\begin{eqnarray}
{\mathbf r.E}^{E,\pm}_{\nu ,m}=-\left(\frac{\mu}{\epsilon}\right)^{\frac{1}{4}}
\frac{\scriptstyle{\sqrt{\nu\left(\nu+1\right)}}}{k}.h^{\left(\pm\right)}_{\nu}\left(\scriptstyle{ kr\sqrt{\epsilon\mu}}\right).Y_{\nu  ,m}\left( \theta,\phi\right)
\nonumber\\
{\mathbf r.H}^{M,\pm}_{\nu ,m}=\left(\frac{\epsilon}{\mu}\right)^{\frac{1}{4}}
\frac{\scriptstyle{\sqrt{\nu\left(\nu+1\right)}}}{k}.h^{\left(\pm\right)}_{\nu}\left(\scriptstyle{ kr\sqrt{\epsilon\mu}}\right).Y_{\nu  ,m}\left( \theta,\phi\right)
\label{cond1}
\end{eqnarray}
and
\begin{eqnarray}
{\mathbf r.H}^{E,\pm}_{\nu ,m}=0  &\quad,\quad&{\mathbf r.E}^{M,\pm}_{\nu ,m}=0\label{cond2}
\end{eqnarray}
In this article we consider only outgoing radiation propagating in vacuum .
 For reasons of generality, we must use two types of indices $\nu_{E}\left(\beta\right)$ and $\nu_{M}\left(\beta\right)$ because of a difference between the boundary conditions for TE and TM waves. For example, if the conductivity of the metallic cone is infinite (the perfect conductor), we must have the boundary conditions:
\begin{eqnarray}
 \left.Y_{\nu_{E} ,m}\left( \theta,\phi\right)\right\vert_{ \theta=\pi-\beta}=0\nonumber\\
\left.\frac{\partial Y_{\nu_{M} ,m}\left( \theta,\phi\right)}
{\partial \theta}\right\vert_{ \theta=\pi-\beta }=0,
 \end{eqnarray}
These expressions fix the authorized $\nu$ values as a function of $\beta$. Hence, an electromagnetic field satisfying specific boundary conditions can be written as an expansion in TE and TM waves, and we have
\begin {eqnarray}
{\mathbf H}=\sum _{\nu_{E},m} a_{\nu_{E},m}^{E}.h^{\left(+\right)}_{\nu_{E} }\left( kr\right)
.\frac{{\mathbf L}Y_{\nu_{E },m}\left( \theta,\phi\right)}{\scriptstyle{\sqrt{\nu_{E} \left( \nu_{E} +1 \right)}}}\nonumber\\-\frac{i}{k}\sum_{\nu_{M},m} a_{\nu_{M},m}^{M}.\nabla \times \left( h^{\left(+\right)}_{\nu_{M} }\left( kr\right)
.\frac{{\mathbf L}Y_{\nu_{M},m}\left( \theta,\phi\right)}{\scriptstyle{\sqrt{\nu_{M} \left( \nu_{M} +1 \right)}}}\right)
\label{sommation2}\end{eqnarray}
and
\begin{eqnarray}
{\mathbf E}=\sum _{\nu_{M},m}  a_{\nu_{M},m}^{M}.h^{\left(+\right)}_{\nu_{M}}\left( kr\right)
.\frac{{\mathbf L}Y_{\nu_{M} ,m}\left( \theta,\phi\right)}{\scriptstyle{\sqrt{\nu_{M} \left( \nu_{M} +1 \right)}}} \nonumber\\
+\frac{i}{k}\sum_{\nu_{E},m}   a_{\nu_{E},m}^{E}.\nabla \times \left( h^{\left(+\right)}_{\nu_{E} }\left( kr\right)
.\frac{{\mathbf L}Y_{\nu_{E} ,m}\left( \theta,\phi\right)}{\scriptstyle{\sqrt{\nu_{E} \left( \nu_{E} +1 \right)}}}\right)
\label{sommation3}\end{eqnarray}
\begin{eqnarray}
{\mathbf E}=\sum _{\lambda} \quad a_{\lambda}\cdot{\mathbf E}_{\lambda}&\quad;\quad&{\mathbf H}=\sum _{\lambda}\quad a_{\lambda}\cdot{\mathbf H}_{\lambda}
\label{sommation}\end{eqnarray}
where $\lambda$ refers to TE and TM waves. In addition it can be demonstrated \cite{Jackson} that we have
 \begin{eqnarray}
a_{\nu,-m}=\left(-1\right)^{m+1}\cdot a_{\nu,m}^{\ast}.\label{moins}
\end{eqnarray}
The coefficients $a_{\lambda}$ are determined by an integration with respect to the solid angle which includes the entire allowed domain $\Omega_{0}$
\begin{eqnarray}
a_{\lambda}=k^{2}\int_{\Omega_{0}}
{\mathbf r}\cdot\left({\mathbf E}\times
{\mathbf H}^{\ast}_{\lambda}\right) r d \Omega.\label{norm}
\end{eqnarray}
Hence, we can express the angular distribution of radiated power, defined as $\frac{dP}{d\Omega}=\frac{c}{8\pi}r^{2}{\mathbf E}\times{\mathbf B}^{\ast}$, as follows:
\begin{eqnarray}
\lefteqn{\frac{d P}{d\Omega}= \frac{c}{8\pi k^{2}}\cdot \|\left(-i\right)^{\nu_{E}+1}
\sum _{\nu_{E},m}  a_{\nu_{E},m}^{\left(E\right)}
.\frac{{\mathbf L}Y_{\nu_{E},m}\left( \theta,\phi\right)}
{\sqrt{\nu_{E}  \left( \nu_{E} +1 \right)}}\times\hat{{\mathbf r}}  } \nonumber \\
 & & {} +\left(-i\right)^{\nu_{M}+1} \sum _{\nu_{M},m}  a_{\nu_{M},m}^{\left(M\right)}
\cdot\frac{{\mathbf L}Y_{\nu_{M} ,m} \left( \theta,\phi\right)}
{\sqrt{\nu_{M} \left( \nu_{M} +1 \right)}}   \|^{2}. \label{poweroflove}
\end{eqnarray}
In addition we can write the total radiated power as:
\begin{eqnarray}
P_{\textrm{total}}=\frac{c}{8\pi}\int_{\Omega_{0}}\textrm{Re}\left(\hat{{\mathbf r}}\cdot\left({\mathbf E}\times {\mathbf H}^{\ast}\right)\right)r^{2}d\Omega\nonumber \\
=\frac{c}{8\pi}\sum_{\lambda}\|a^{+}_{\lambda}\|^{2}.
\end{eqnarray}
%
%
\section{The radiation pattern of a coated tapered fiber tip}
\begin{figure}[h]
\begin{center}
\includegraphics[width=10cm]{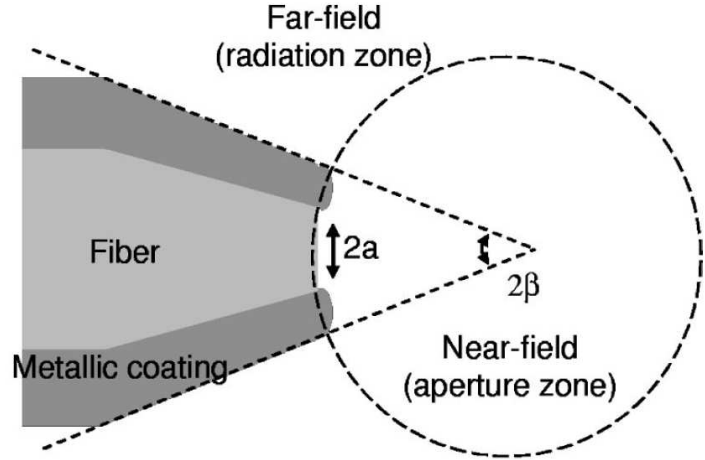}
\caption{An idealized metal-coated fiber tip with its characteristic funnel shape. The top angle $\beta$ is inferior to $15^{\circ}$, and the aperture diameter $2a$ is of the same order of magnitude as the aluminum coating thickness, i.~e., 100 nm.  }
\label{schema2}
\end{center}
\end{figure}

The tapered and metal-coated optical fiber tip used in NSOM\cite{Synge,Betzig,Pohl} is characterized by a ``funnel'' geometry in which an aluminum layer (thickness of $\sim 100$~nm) is evaporated onto a glass cone truncated at its apex by a small disk (diameter of $2a\sim 60 $ nm, see Fig.~\ref{schema2}).

\begin{figure}[h]
\begin{center}
\includegraphics[width=10cm]{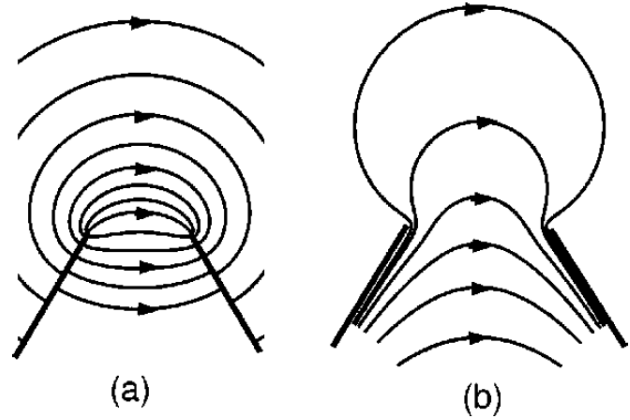}
\caption{ The electric (a) and magnetic (b) fields in the near-field zone of a two-dimensional cone, calculated rigourously with adapted complex potentials. The assumption is made that the topology of the 3D problem can be obtained by  deforming the 2D case. In the 3D case, (a) and (b) planes are perpendicular to each other. In addition, the magnetic field  in (a) and  the electric field in (b) are perpendicular to both planes for symmetry reasons.     }
\label{schema4}
\end{center}
\end{figure}

The shape of the tip reveals a partially conical geometry. Therefore, the system is -to first approximation- equivalent to a conical antenna and it can be described by a quasi-multipole expansion.
Nevertheless, in order to take into account the finite size of the aperture domain in our modelization, we  study the field outside a sphere of radius $a/\sin{\beta}$ only. This so-called aperture zone represents the ``terra incognita'' near-field  region ($r/\lambda\leq 1$) of the fiber tip (see Fig.~\ref{schema2}).
Let $\epsilon\simeq\mu\simeq1$ be the electric and magnetic permittivities of vacuum. The electromagnetic field in the region $r/\lambda\geq 1 $ can be described by an expansion using the quasi-multipole formalism.
  The coefficients $a_{\nu,m}$ contain all properties of the electromagnetic field.\\
 In order to calculate ${\mathbf E}$ and ${\mathbf B}$ in the radiation zone, we need (see Eq.~\ref{norm}) to know the field on the spherical section $\Omega_{0}$, i.~e. in the near-field zone of the aperture. Unfortunately, there is as yet no complete theory of light diffraction by a small aperture in a screen with conical geometry, and we must rely on approximations.
The $a_{\nu,m}$ coefficients can be evaluated by using a Taylor series written in the vicinity of the tip origin. We obtain, using Eq.~\ref{norm}, the following expressions to  first order in ${\mathbf r}$:
\begin{eqnarray}
a_{\lambda}\simeq 2\pi i k^{3}\{-{\mathbf E}\left(0\right)\cdot\frac{1}{4\pi}\int_{\Omega_{0}}
{\mathbf r} \times\left(\hat{{\mathbf r}}\times {\mathbf E}_{\lambda}^{\ast}\right)\ r^{2} d \Omega  \nonumber\\
 + {\mathbf B}\left(0\right)\cdot \frac{1}{2 i \pi k }\int_{\Omega_{0}}\left(\hat{{\mathbf r}}\times {\mathbf E}_{\lambda}^{\ast}\right) r^{2} d \Omega  + ...\}\nonumber \\
+\frac{1}{6}\sum_{ \alpha,\beta}[ \mathcal{Q}_{\lambda,\alpha,\beta}\partial_{\beta}E_{\alpha}\left(0\right)-\mathcal{M}_{\lambda,\alpha,\beta}\partial_{\beta}B_{ \alpha}\left(0\right)].
\label{truc}
\end{eqnarray}
In Eq.~\ref{truc}, the symmetric terms with prefactors $\mathcal{Q}_{\lambda,\alpha,\beta}$ and $\mathcal{M}_{\lambda,\alpha,\beta}$ are neglected in a first approximation. The integrals can be evaluated using the following properties: i) the $\beta$ value is small and, as a consequence, the fraction $\left(4\pi-\Omega_{0}\right)/\left(4\pi\right)$ is negligible, ii) the authorized $\nu$ values are close to 1, and iii) there are only the two first roots $\nu_{E,p=1}$, $\nu_{M,p=1}$ associated with our boundary problem. These second and third hypothesis will be justified later on but they can be intuitively understood by realizing that in the limit of  a very small tip angle $\beta $, the theory must reduce to the case of a linear antenna in which the first and dominating radiation mode is the dipole term with $l=1$. Hence, using Maxwell's equations, we have
\begin{eqnarray}
a^{E}_{\nu_{E},m}\simeq -2\pi i k^{3}\langle{\mathbf E}_{\nu_{E} ,m}^{ E \ast}\rangle.
\frac{1}{3}\frac{a^{3}}{\left(\sin{\beta}\right)^{3}}{\mathbf E}\left(0\right)  \nonumber \\
a^{M}_{\nu_{M},m}\simeq -2\pi i k^{3}\langle{\mathbf B}_{\nu_{M},m}^{M \ast}\rangle.
\frac{2}{3}\frac{a^{3}}{\left(\sin{\beta}\right)^{3}}{\mathbf B}\left(0\right),
\label{expan}\end{eqnarray}
where the average is taken in the spherical domain of radius $a/\sin{\beta}$.\\
It can be noted that in the spherical harmonic expansion, the terms $\|m\|\geq 1$ are prohibited by recursion relations, but this is not the case here.\\ In order to calculate $a_{\lambda}$, we need to know the local field ${\mathbf B}\left(0\right),{\mathbf E}\left(0
\right)$ at the center of the aperture. We would like to mention that a rigorous representation of the near-field can be drawn for a two-dimensional tip as shown in Fig.~\ref{schema4}. Such a tip is described by magnetostatic and electrostatic potentials which can be found using the well-known complex potential method valid for the Laplace equation in two dimensions \cite{Durand}. Numerical calculations by Novotny et al~.
\cite{Novotny94} shows the same field behavior. The field topology in the 3D case can be anticipated from this 2D result. The choice of this field topology is based on the assumption that the mode
entering the conical part of the tip is the one in the preceeding hollow, circular cylinder. This is principally the polarized TE11 mode used in near-field microscopy.\\ Turning back to the 3D case, we remark that, despite the lack of a rigorous analytical model, the near-field can be computed numerically using different methods. These calculations confirm the intuitive field topology obtained in our 2D result. Using our 2D model and numerical computations obtained by other authors, we postulate that the unknown near-field components  are linked to each other by the ``plane wave'' relation :
\begin{equation}
\langle{\mathbf E}\rangle=\langle{\mathbf B}\rangle\times\hat{{\mathbf z}}, \label{waveplane}
\end{equation}
where $z$  is the cone symmetry axis. This condition may be intuitively understood by assuming that the incident light propagating in the fiber is linearly polarized. This means that the symmetry of the field is conserved at the aperture as shown in Fig.~\ref{schema4}.
The electric and magnetic amplitudes at the center of the aperture domain (i.~e. at the top of the cone) can be determined from the Poynting theorem\cite{Landau} which imposes the equality $\int_{V} \| {\mathbf B}\|^{2}= \int_{V} \| {\mathbf E}\|^{2}-i\frac{c}{\omega}\oint\left({\mathbf E}\times {\mathbf B}^{\ast}\right){\mathbf r}rd\Omega$ in the spherical domain of the aperture zone. If the surface integral is equal to zero in this formula, i.~e. if most of the energy is either transmitted or reflected  but not stocked or dissipated in the near-field zone, we obtain $E\left(0\right)\simeq B\left(0\right)$ as a good approximation.\\
Hence, with this symmetry condition, the field is entirely determined and the calculation shows that the only
non-vanishing terms have $m=1$. The normalized  angular distribution of radiated power is then
  \begin{equation}
\displaystyle
\frac{1}{P}\frac{d P}{d\Omega}\simeq\frac{2}{5}\left\|
\textrm{Im}\frac{{\mathbf L}Y_
{\nu_{E} ,1}\left( \theta,\phi\right)}{\sqrt{\nu_{E} \left( \nu_{E} +1 \right)}}
  \times \hat{{\mathbf r}} +2\textrm{Re}\frac{{\mathbf L}Y_{\nu_{M} ,1}\left( \theta,\phi\right)}
{\sqrt{\nu_{M} \left( \nu_{M} +1 \right)}} \right\|^{2},
\label{epsilon}
\end{equation}
in which only the $\nu$ values $\nu_{E}$ and $\nu_{M}$ have been considered .\\
It is worth noting here that our derivation of Eq.~\ref{epsilon} can be obtained in a similar way by means of a quasi spherical eigenfunction expansion of the scalar Green's functions for a cone\cite{Felsen1,Felsen2,Uslenghi,ChenToTai,Siegel} with  Dirichlet (or Neumann type) boundary conditions depending on $\nu_{E}$ (respectively $\nu_{M}$) and $m$. The scalar field $\bbox{r}\cdot\bbox{E}$ (respectively $\bbox{r}\cdot\bbox{B}$ ), and consequently the  coefficients $a^{E}_{\nu_{E},m}$ (respectively $a^{M}_{\nu_{M},m}$ ) given by Eq.~\ref{expan} can be easily deduced from a surface integral on $\Omega_{0}$. Conical Green's functions (which have already been used in the context of near-field optics for aperturless microscopes\cite{Cory,Aigouy}) connect surface integrals, given by Eq.~\ref{norm} and Eq.~\ref{expan}, to the Huygens-Fresnel principle directly. \\In order to complete this section we can note than the ratio $\frac{1}{P}\frac{d P}{d\Omega} $ represents the physical quantity in measurements of the angular power emitted by actual fiber tips as explained later on. The total radiated power of the fiber tip can be expressed as a function of the field ${\mathbf E}\left(0\right)$
\begin{equation}
P_{\textrm{total}}\simeq  \frac{5c}{864 k^{2}}\left(\frac{ka}{\sin{\beta}}\right)^{6}{\mathbf E}\left(0\right)^{2}.
\end{equation}
 This formula can be compared with the Bethe\cite{Landau} result
$P=\frac{5c k^{6}a^{6}}{108\pi^{2}k^{2}}{\mathbf E}_{0}^{2}$, where $E_{0}$ is the field at the center of the aperture. The same behavior appears in the two formulas: a dependence in $a^{6}$ and in $E_{0}^{2}$. The difference arises from the
$\left(\sin{\beta}\right)^{6}$ which represents the second geometrical parameter of the fiber tip. It is important to note that this formula contains an unknown variable: the field $E_{0}$. This field is a function of the transmission of the tip which depends on the over-exponential decay in the conical wave guide due to the cut-off of the propagating mode in the fiber\cite{Novotny}. This decay is not taken into account in our model which prevents us from computing the tip transmission.
%
%
\section{ Boundary conditions and perturbation method}
 In order to complete the solution of Maxwell's equations, we will describe the boundary conditions on the cone and justify the above assumption $\nu\simeq 1$. A simple analogy  with wave guide theory permits to establish the boundary conditions in a conical geometry (this  corresponds to the case of a ``wave guide'' of variable cross section). For a perfect metal , we  have
\begin{eqnarray}
\left. Y_{\nu_{E} ,m}\left( \theta,\phi\right)\right\vert_{S}=0,&\left.
\frac{\partial Y_{\nu_{M} ,m}\left( \theta,\phi\right)}
{\partial \cos{\theta}}\right\vert_{S}=0\label{cond}
 \end{eqnarray}
These equations impose a condition on $\nu_{E}$ and $\nu_{M}$ which restricts the allowed values to infinite, growing sequences $\nu_{E,p}$ and $\nu_{M,p}$ that depend on $\beta$ and $m$ (p integer). Because $R=\frac{ka}{\sin{\beta}}\leq 1$ in the near-field, two successive roots $\nu' \leq \nu$ associated with a TM mode (or a TE mode, respectively) obey:
\begin{equation}
\frac{a_{\nu', m'}^{E, M}}{a_{\nu, m}^{E,M}}\sim \left(kR\right)^{\nu'- \nu} \ll 1.
\end{equation}
 This allows to neglect, to a good approximation, all roots except for the first two with $p=0$. The calculation of these roots is possible for a small cone angle $\beta$ using the approximation formula of Hobson and Schelkunoff\cite{Hall,schelkunoff} which give for $m=1$:
\begin{eqnarray}
\nu_{E,0,m=1} \simeq 1+\frac{1}{2\ln{\frac{2}{\beta}}+\frac{2}{\left(\beta\right)^{2}}}\nonumber\\
\nu_{M,0,m=1} \simeq 1+\frac{1}{2\ln{\frac{2}{\beta}}+\frac{6}{\left(\beta\right)^{2}}}.
\end{eqnarray}
For a tip angle of $15^{\circ}$ we have $\nu_{0}\simeq 1.03$. This approximation can be compared with the values obtained by solving Eqs.~\ref{cond} numerically. Here we have for the same $\beta$ $\nu_{E,0}\simeq 1.033$ and $\nu_{M,0}\simeq 0.967$. \\
In reality, aluminum is a good but not perfect metal which possesses a frequency-dependent complex dielectric constant $\epsilon_{c}=\epsilon_{c}'+i\epsilon_{c}''$. In the optical domain for $\lambda\leq 800$ nm, the permittivity of aluminum can be described by the Lorentz formula:
 \begin{equation}
\epsilon_{c}\left(\omega\right)\simeq 1-\frac{\omega_{p}^{2}}{\omega^{2}+i\omega\gamma}.
\end{equation}
The plasma frequency and the damping constant are given\cite{Novotny} by $\omega_{p} \simeq 15.56 \textrm{eV}$ and $ \gamma\simeq 0.608 \textrm{eV}$, respectively. At $\lambda\simeq 633$ nm, we have $\epsilon_{c}\simeq -54.4 +i18.8$  which implies a refractive index $n_{c}=\sqrt{\epsilon}=1.18+i7.12$ and a skin depth
$\delta=\frac{1}{2 k n''_{c}}\simeq 7 \textrm{nm}$ which is small but not negligible compared to the metallic clading of $\simeq 100$ nm.  The influence of the finite skin depth on the ``perfect'' and unperturbed TM and TE modes  may be taken into account.\\ The perturbed  solution is discussed in the textbook of Jackson\cite{Jackson}. The result is a condition linking ${\mathbf E}$ and ${\mathbf B} $ on the surface with the aluminum permittivity \cite{Jackson}:
\begin{equation}
{\mathbf E}_{\|}= {\mathbf E}- \hat{{\mathbf \theta}}  \left({\mathbf E}\cdot\hat{{\mathbf \theta}} \right) \simeq-\sqrt{\frac{1}{\epsilon_{c}}}\hat{{\mathbf \theta}}  \times {\mathbf B}.\label{limit}
\end{equation}
 This implies an effective current confined to the surface obtained by integration of Ohm's law in the aluminum layer of thickness $\delta$ located just below the conductor surface.  Using Eq.~\ref{limit}, we can establish
\begin{eqnarray} \left.Y_{\nu_{E} ,m}\right\vert_{S}= -i\sqrt{\case{1}{\epsilon_{c}}}\case{kr}{\nu_{E}\left(\nu_{E}+1\right)}\left.\case{\partial Y_{\nu_{E} ,m}}{\partial \theta}\right\vert_{S}
\nonumber \\
\left.\case{\partial Y_{\nu_{M} ,m}}{\partial \theta}\right\vert_{S}= i\sqrt{\case{1}{\epsilon_{c}}}\case{\nu_{M}\left(\nu_{M}+1\right)}{kr}\left. Y_{\nu_{M} ,m}\right\vert_{S}. \label{condpert}
 \end{eqnarray}
As expected, this reduces to the unperturbed condition for $\sigma=\infty$. By using the substitution $\nu \rightarrow \nu^{\left(0\right)}$ on the right-hand side of Eq.~\ref{condpert} we have to lowest order $ \psi_{E}\vert_{S}\simeq f_{E}\frac{\partial \psi^{\left(0\right)}_{E}}{\partial n}\vert_{S} $ and $\frac{\partial \psi_{M}}{\partial n}\vert_{S}\simeq f_{M}\psi^{\left(0\right)}_{M}\vert_{S}$ where the factors $\|f_{E,M}\|$ are small.
In order to calculate the perturbed $\nu$, we use the two-dimensional Green's theorem in its integral form\cite{Jackson} and we obtain:
\begin{eqnarray}
\nu^{\left(0\right)}_{E}\left(\nu^{\left(0\right)}_{E}+1\right) - \nu_{E}\left(\nu_{E}+1\right)  \simeq \left(1.4+i 0.23\right)kr
\nonumber\\
\nu^{\left(0\right)}_{M}\left(\nu^{\left(0\right)}_{M}+1\right) - \nu_{M}\left(\nu_{M}+1\right) \simeq \left(1.4+i 0.23\right)10^{-5}\frac{1}{kr}.
\end{eqnarray}
 $\nu_{E,0}$ and $\nu_{M,0}$ are functions of two radii $r_{E}\simeq 18$ nm and $r_{M}= \lambda =633$ nm that are adjusted to reproduce the experimental results discussed below. The two values of $\nu$ are complex and the real parts are close to unity as expected: $\nu_{E}\simeq 0.95-0.01i$, $\nu_{M}\simeq 0.96$. These fits reveal that the adapted boundary conditions depend on a typical length $r<\lambda$ that is probably linked to the aperture radius $a\simeq 20 nm$ (this value is in agreement with the aperture radius obtained by scanning electron microscopy\cite{Karraia}). With these values of $\nu$ we can calculate the radiated power distribution of the tip\cite{Drezet}.
\section{Comparison with experimental data}
\begin{figure}[h]
\begin{center}
\includegraphics[width=10cm]{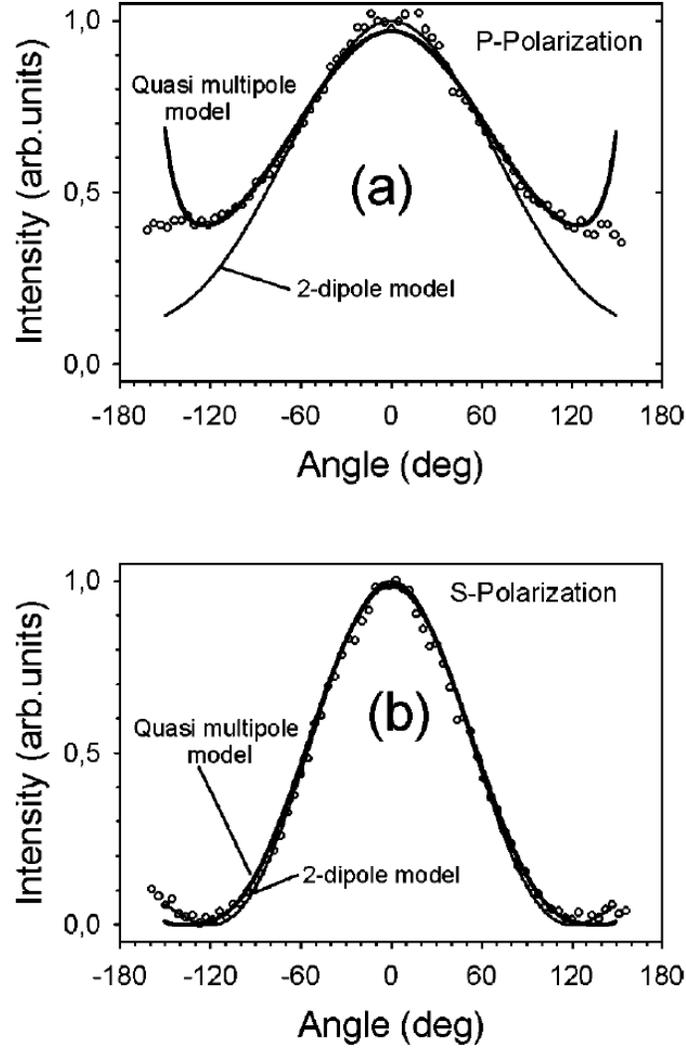}
\caption{``Orbital'' representation of the angular radiated power of the tip
($r=I\left( \theta,\phi\right),\theta,\phi$ ). The figure shows the important backward emission in the P plane passing through the tip axis and containing the effective electric dipole ${\mathbf P}$.}
\label{article3rad}
\end{center}
\end{figure}

\begin{figure}[h]
\begin{center}
\includegraphics[width=10cm]{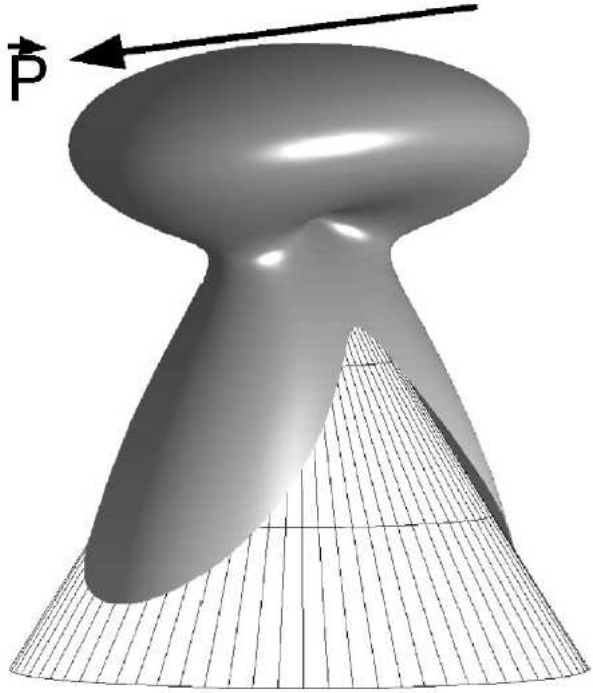}
\caption{ 3 D isoline representation of the electromagnetic energy density (arb. units) in the far-field zone
  $kr\gg 1$. This figure shows the difference in the far-field energy distribution that exists between the S and the P polarization planes, and the presence of an important backscattering effect in the P plane. Additionally, the tip of the metallic cone is drawn in blue. The sphere at the tip apex corresponds to the (unknown) near-fied zone where the far-field divergent energy distribution $\propto \case{1}{r^{2}}$ is no longer valid.}
\label{article3ff}
\end{center}
\end{figure}

\begin{figure}[h]
\begin{center}
\includegraphics[width=10cm]{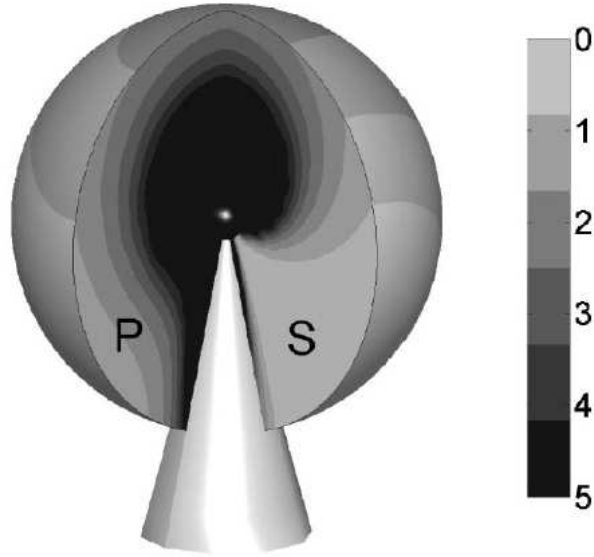}
\caption{Normalized angular distribution of radiated power for a) the P polarization where the incident electric field is parallel to the analysis plane, and b) for the S polarization where the detector is scanned
perpendicularly to the plane of polarization of the incident light. Experimental data are shown in circles. The thick line corresponds to the present quasi-multipole model, the thin line to the 2-dipole model. The experimental data (aperture radius $a=20$ nm, light  wavelength $\lambda=633$ nm) are taken from Refs.~1,2.}
\label{SandP}
\end{center}
\end{figure}

Measurements of the far-field radiated power of tapered fiber tips in S and P plane have been carried out extensively by Oberm\"uller and Karrai \cite{Karraia,Karraib} . As seen in Fig.~\ref{SandP}, there is an important backscattering effect in the P plane. The experimental data and the results obtained with the quasi-dipole model (using the values of $\nu_{E,M}$ obtained above) are also shown in Fig.~\ref{SandP} for comparison\cite{Drezet}. The agreement is good for most angles with the exception of extreme $\theta$ values corresponding to grazing observation angles on the metallic cone. The disagreement between theory and experiment is probably due to higher terms in the expansion that we have neglected. This is confirmed by the recent independent results obtained by Shin et al.\cite{Shin} which have fitted the experiment to an expansion in classical spherical multipoles. Their fit contains dipolar terms (where the factor $2$ plays the same role as in our model) as well as quadrupolar ($l=2$) and octopolar ($l=3$) terms. The behavior of the field in the extreme regions probably also depends on a modified diffusion due to surface roughness\cite{Karraiprivate}.\\
The dipole model of Oberm\"uller and Karrai, which corresponds to $\nu=1$ (a value close to the real values $\nu_{E,0}$ and $\nu_{M,0}$), gives good results for the S polarization but cannot reproduce the backward emission in the P plane.
 The fact that the two values of $\nu_{E,M}$ approach unity explains the efficiency of the dipole model for small azimuthal angles ( in this regime $Y_{\nu,m}\simeq Y_{1,m}$). In addition we can formally justify the dipole model considering Eq.~\ref{expan} as the generalization for conical geometry of the fundamental definition of multipole moments which enter into the spherical harmonic multipole expansion, and which are used for localized source distributions. Using this analogy, we deduce the two effective dipoles associated with the aperture
\begin{eqnarray}
{\mathbf P}_{\textrm{eff}}=\frac{1}{4\sqrt{2}}\frac{a^{3}}{\sin{\beta}^{3}}{\mathbf E}\left(0\right)\nonumber \\
{\mathbf M}_{\textrm{eff}}=2\hat{{\mathbf z}}\times{\mathbf P}_{\textrm{eff}}=2\frac{1}{4\sqrt{2}}\frac{a^{3}}{\sin{\beta}^{3}}{\mathbf B}\left(0\right)\label{dipoles}.
\end{eqnarray}
Once again, the characteristic factor of 2 between the magnetic and electric moments appears in this formula.
Figs.~\ref{article3rad} and \ref{article3ff} are two equivalent representations of the spatial dependence of radiated power. In particular, the ``orbital'' representation of Fig.~\ref{article3rad} shows the important backward emission in the plane containing the effective electric dipole ${\mathbf P}$ and the tip axis (P plane). This backward effect is less important in the S plane containing the tip axis and the magnetic dipole due to the strong near-field effect on the metal surface.
 The factor of two in Eq.~\ref{dipoles} (as in Eq.~\ref{epsilon}) stresses the difference between electric and magnetic dipoles as discussed in the introduction (namely an oscillating charge in the case of an electric dipole and a rotating charge for a magnetic dipole). This factor of two and the dependence of the dipoles on $a$ and  on $E_{0}$ are similar to the Bethe model and the Rayleigh diffusion theory by subwavelength particles and holes, but the dipole orientations  are different, a result unexplained by the Bethe-Bouwkamp solution. \\We can note in addition that the far fields radiated by a small distribution of electric and magnetic dipoles located in the apex zone which satisfy a relation of the type given by Eq.~\ref{dipoles}, are equivalent to those deduced from Eq.~\ref{expan} (and consequently Eq.~\ref{epsilon}) in the limit of subwavelength distributions. This fact, which can be obtained by means of  scalar Green's functions discussed above, gives further support to the analogy between dipoles and aperture at the apex of a cone. \\
It is worthwhile to mention  that A.~Roberts\cite{Roberts} has used the method of quasi-multipoles for the interior problem of ``small hole coupling of radiation into a near field probe''. In this   problem the author used the limit of small $\beta$ where $Y_{\nu,m}\left(\theta, \phi\right)\sim J_{\nu}[\left(\nu+\case{1}{2}\right)\theta]e^{im\phi}$ and showed that the coupling of a wave into a conical wave guide implies effective dipoles given by the Bethe-Bouwkamp solution. These dipoles are perpendicular to the aperture, a result which is completely different from the problem of radiation by a tip which generates dipoles located in the aperture plane.\\
For completeness, we wish to comment on some additional, interesting properties of the emission by conical tips that are relevant to the experiment. The quasi-multipole model shows that the aperture radius disappears in the formula Eq.~\ref{epsilon} and, consequently, in the radiation profile. We can understand this fact using the analogy with Rayleigh's and Mie's theory\cite{Born} of diffusion by small particles which follow the same behavior as our model and which can be associated with the impossibility to see details smaller than the wavelength of light. In particular, the aperture size cannot be a relevant variable in the limit of small apertures. For the same reason, the tip angle $\beta$ does not play an important role in the diffraction profile. The dependence on the aperture radius is very different in the short-wavelength limit and for large apertures $a\geq\lambda$.
For a large hole, the transmission must approach unity and, by consequence, the aperture radius may  appear in the profile . We have then
$\frac{dP}{Pd\Omega}\sim \left(ka\right)^{2}\cos^{2}\left(\theta\right)\left(\frac{2J_{1} \left(ka sin\left(\theta\right)\right)}{ka \sin\left(\theta\right)}\right)^{2}$ and $P\sim c E_{0}^{2}a^{2}/8$ that can be compared with
$P\sim\frac{c k^{6}a^{6}}{k^{2}}{\mathbf E}_{0}^{2}$, valid for Bethe's theory and our conical subwavelength aperture model .
%
%
%
\section{summary}
An optical fiber tip with a small aperture in the metal coating as used in NSOM is usually characterized by its angular transmission profile. In order to describe this profile theoretically, we have presented in this paper an analytical model for light diffraction through a subwavelength aperture located at the apex of a metallic screen with conical geometry. For this purpose, a field expansion in quasi spherical multipoles has been developed to solve Maxwell's equations analytically. Special care has been taken to consider realistic boundary conditions adapted not only to the conical geometry but also to the finite conductivity of the coating.
Our model is able to reproduce and
to explain the experimental results. The diffraction in the far-field is demonstrated to remain similar to that drawn with the 2-dipole model\cite{Karraia,Karraib} over a large portion of space. The latter model is formally justified in our approach. In particular, we find a characteristic factor of 2 between the expressions for the magnetic and the electric dipole (which is reminescent of the Rayleigh diffusion theory by small particles and the Bethe problem).
In addition, the previously unexplained large ``backward'' emission in the P plane (which does not exist in the S plane) results directly from our analysis. This ``P emission'' arises from a Poynting vector flow along the coated surface of the tip, an effect that cannot be described within a simple dipolar model.
%


\newpage

All captions are in the order of appearence in the text.

\end{document}